\documentclass[a4paper]{article}

\usepackage[colorlinks,allcolors=red]{hyperref}
\usepackage{INTERSPEECH2019}
\usepackage{graphicx}
\usepackage{enumitem}
\usepackage{booktabs}
\usepackage{multirow}
\usepackage{amsmath}

\title{Few Shot Speaker Recognition using Deep Neural Networks}
\name{Prashant Anand$^1$, Ajeet Kumar Singh$^1$, Siddharth Srivastava$^{1,2}$, Brejesh Lall$^1$}

 \address{
   $^1$Indian Institute of Technology Delhi, India\\
   $^2$Centre for Development of Advanced Computing, Noida}
 \email{prs0704@gmail.com, ajeetsngh24@gmail.com, siddharthsrivastava@cdac.in, brejesh@ee.iitd.ac.in}

\begin{document}

\maketitle
\begin{abstract}
The recent advances in deep learning are mostly driven by availability of large amount of training data. However, availability of such data is not always possible for specific tasks such as speaker recognition where collection of large amount of data is not  possible in practical scenarios. Therefore, in this paper, we propose to identify speakers by learning from only a few training examples. To achieve this, we use a deep neural network with prototypical loss where the input to the network is a spectrogram. For output, we project the class feature vectors into a common embedding space, followed by classification. Further, we show the effectiveness of capsule net in a few shot learning setting. To this end, we utilize an auto-encoder to learn generalized feature embeddings from class-specific embeddings obtained from capsule network. We provide exhaustive experiments on publicly available datasets and competitive baselines, demonstrating the superiority and generalization ability of the proposed few shot learning pipelines.

\end{abstract}

\section{Introduction}
Speaker recognition with only a few and short utterances is a challenging problem. In this paper, we assume that for each speaker, only a few very short utterances are available. Specifically, we learn from $1$ to $5$ utterances of $3$ seconds each for each speaker unlike earlier work where recordings  upto $10$ seconds are considered as a short utterance. Apart from computational and technical improvisations obtained by solving under such constraints, this setting also allows enrolment of speakers to be easier and more practical. 

Recently many approaches have been proposed for speaker identification using deep neural networks~\cite{snyder2018x, snyder2017deep, lei2014novel} but they assume availability of large amount of training data. Moreover, the deep learning pipelines attempting to learn from short utterances~\cite{guo2018deep} are, in general, based on i-vectors or Mel-frequency cepstral coefficients (MFCCs). While MFCCs are known for susceptibility to noise~\cite{zhao2013analyzing}, the performance of i-vectors tend to suffer in case of short utterances~\cite{kanagasundaram2011vector}. Moreover, it has been shown that convolutional neural networks (CNNs) are able to mitigate the noise susceptibility of i-vectors, MFCCs~\cite{mclaren2014application} and have been successfully used for speaker recognition~ \cite{liu2018gmm}. Since convolutional neural networks are data-hungry, and are able to exploit structured information such as images very effectively, recently large scale speaker recognition datasets  have been made publicly available where benchmarks setup on CNNs with spectrogram as input have been shown to perform very well~\cite{Nagrani17, Chung18b}. However, their effectiveness to learn or generalize with limited amount of data (few-shots)
 and short utterances is not established very well.
 
Few shot learning paradigms have recently been effectively applied for audio processing~\cite{chou2018learning, arik2018neural}. However, their effectiveness to speech processing especially speaker recognition is still unknown. To this end, we utilize CNN as base network with spectrogram derived directly from raw audio files as input and evaluate the effectiveness of these networks in case of constrained setting of few shot learning for speaker identification. We choose VGGNet~\cite{Simonyan14c} and ResNet~\cite{dr} as the base architectures and evaluate them under various settings. For generalizing them under unseen speakers, we use prototypical loss~\cite{NIPS2017_6996}. 

While CNNs have shown to perform very well, but they are not able to exploit the spatial relationships within an image. Bae et al~\cite{bae2018end} argued that CNNs are not able to leverage the spatial information within spectrograms such as between pitch and formant. Therefore, they utilized capsule networks~\cite{sabour2017dynamic} for speech command recognition. Based on their work, we argue that exploiting spatial relationships in spectrogram images can lead to better speaker recognition as well. However, their are two problems in applying capsule network for speaker recognition. First is that their applicability to complex data is yet to be established~\cite{xi2017capsule} and hence the generalization ability. Second, they are extremely computationally intensive. We reduce the computational complexity by dropping the reconstruction loss from the default capsule network. Next, we add an auto-encoder to map the class feature vectors from capsule network to a common embedding space. The projected feature vector from the embedding space are then subjected to prototypical loss to learn from the constrained data. The entire pipeline is trained end-to-end. 

In view of the above, following are the contributions of this paper
\begin{itemize}[leftmargin=*]
\item To the best of our knowledge, this is the first work that poses speaker recognition as a few-shot learning problem and applies convolutional neural networks and capsule network for speaker recognition under the constraints of short and limited number of utterances. 
\item We show that using convolutional neural network having spectrogram as input and prototypical loss, a speaker can be identified with high confidence with only a few training samples and very short utterances ($\sim3$ seconds).  
\item We propose a novel network based on Capsule Network by significantly reducing the number of parameters and learning a class embedding using auto-encoder with a prototypical loss for generalizing capsule network to unseen speakers. We show that the proposed method performs better than VGG with equivalent number parameters while lags behind ResNet having significantly higher number of parameters. 
\item We perform exhaustive experiments on publicly available datasets and analyze the performance of the considered networks under various settings.
\end{itemize}
\begin{figure*}
    \centering
    \includegraphics[width=12.8cm]{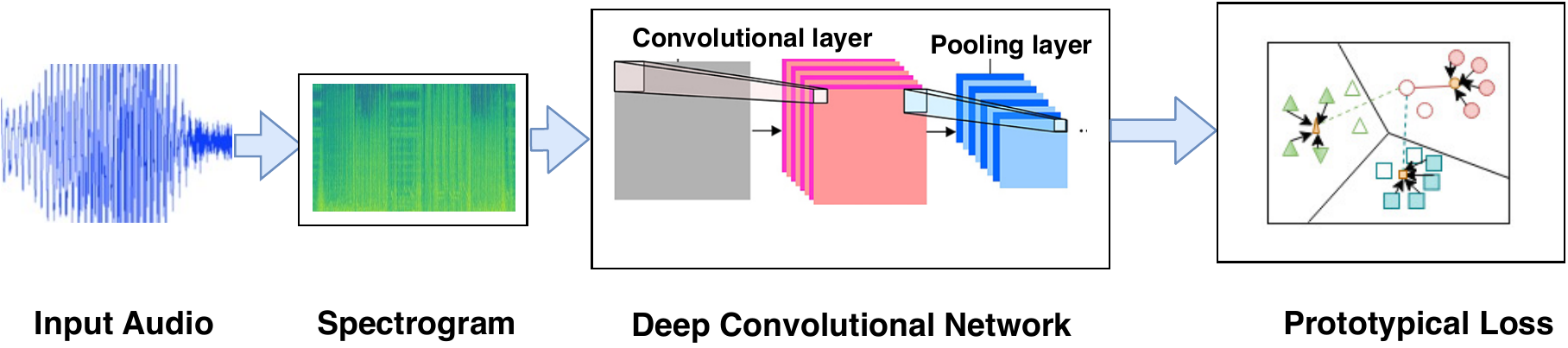}
    \caption{Flow diagram for few shot learning with deep neural networks.}
    \label{fig:flow}
\end{figure*}
The rest of the paper is organized as follows. Section~\ref{sec:methodology}, discusses the various steps and the proposed pipeline. In Section~\ref{sec:exp}, results are discussed while in Section~\ref{sec:conc}, conclusion is provided.

\section{Methodology}\label{sec:methodology}

Figure~\ref{fig:flow} shows the pipeline for using spectrogram with deep neural networks. The input of the network is a spectrogram obtained from raw audio file. For few shot learning, the network may be pre-trained with a large dataset. Now the task is to classify new speakers for whom, we have limited number of samples. Here we additionally, pose a constraint that along with limited number of samples, the duration of each sample is limited to $3$ seconds only. For classification, the existing network is fine-tuned with the training samples of the new speakers. However, as demonstrated in experiments, directly using the embeddings obtained from a pre-trained network causes significant drop in the performance. Therefore, we propose to use prototypical loss to optimize the embeddings by forming representative prototypes (Section~\ref{sec:proto}) . 

We have evaluated against two types of networks~\textit{viz.} Convolutional Neural Networks (VGG, ResNet) and Capsule Networks. While extending CNNs for prototypical loss is straight-forward as the provide class agnostic feature embeddings and hence the prototypes can be learned directly from the feature vectors. However, it is not the case with Capsule Networks as they learn class specific embeddings. Therefore, we learn a generalized embedding using an auto-encoder prior to using prototypical loss (Section~\ref{sec:autocaps}). We now describe each component in detail.

\subsection{Spectrogram Construction}
\label{spectrogram}
First we convert all audio to single-channel, $16$-bit streams at a $16$ kHz sampling rate for consistency. The spectrograms are then generated by sliding window protocol by using a hamming window. The width of the hamming window is $25$ ms with a step size of $10$ ms. This gives spectrograms of size $128$ (number of fft features) x $300$ for $3$ seconds of randomly sampled speech for each audio.  Subsequently, each frequency bin is normalized (mean, variance). the spectrograms are constructed from the raw audio input i.e. no pre-processing such as silence removal etc. is performed. 

\vspace{-0.7em}

\subsection{Model}

\begin{figure*}
    \centering
    \includegraphics[width=15.8cm]{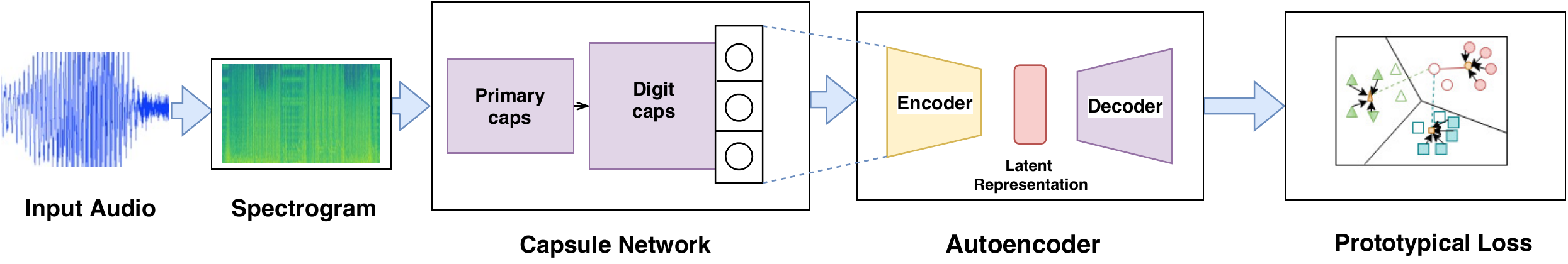}
    \caption{Flow diagram for few shot learning with Capsule Networks.}
    \label{fig:flow1}
\end{figure*}

\subsubsection{Capsule Network}
Capsule network proposed by Hinton~\cite{sabour2017dynamic} replaces the scalar output neuron present in Deep Neural Networks (DNNs) and Convolutional Neural Networks (CNNs) with a capsule (a vector output neuron). This enables capsule network to capture pose information and contain it in a vector. The network is trained by "dynamic routing between capsules" algorithm. Capsule Network has a convolutional layer with stride $1$, on a primary capsule layer and a dense capsule layer. The convolution operation in primary capsule layer has stride $2$. Our modified capusle network, CapsuleNet-M has sride $6$ in both - the first convolution layer and primary capsule layer. Capsule Network is trained with margin loss shown below.

\begin{equation}
\begin{split}
L_{c}=T_{c} \max (0,& m^{+}-\left\|\boldsymbol{v}_{c}\right\| )^{2} \\ &+\lambda\left(1-T_{c}\right) \max \left(0,\left\|\boldsymbol{v}_{c}\right\|-m^{-}\right)^{2}
\end{split}
\end{equation}

Here, $L_{c}$ is the margin loss of class $c$, $v_{c}$ is a final output capsule in class $c$, $T_{c} = 1$ iff the target class is $c$, $m^{+} = 0.9$, $m^{-} = 0.1$ and $\lambda = 0.5$.

\subsubsection{VGG-M and Resnet-34}
VGG and Resnet are Deep CNN architecture which have very good classification performance on image data, so, we chose to use these models with slight modifications to adapt to spectrogram input. These modeified VGG and Resnet-34 models were introduced in~\cite{Nagrani17} and~\cite{Chung18b} respectively.  The architectures of VGG-M amd Resnet-34 used are specified in Table~\ref{vgg_arch} and~\ref{res_arch} respectively.

\vspace{-0.7em}
\subsection{Few shot learning using Prototypical Loss}\label{sec:proto}

In few shot learning, at test time we have to classify test samples among $K$ new classes given very few labeled examples of each class. We are provided with $D$ dimensional embeddings $\mathbf{x}_{i} \in \mathbb{R}^{D}$ for each input $\mathbf{x}_{i}$ and its corresponding label $y_{i}$ where $y_{i} \in\{1, \ldots, K\}$. The objective is to compute a $M$ dimensional representation i.e. the prototype of each class $\mathbf{a}_{k} \in \mathbb{R}^{M}$. The embedding is computed via a function $f_{\phi} : \mathbb{R}^{D} \rightarrow \mathbb{R}^{M}$ where $f$ indicates a deep neural network and $\phi$ indicates its parameters. The prototype $\mathbf{a}_{k}$ is the mean of the support points for a class. 

By using a distance function $d$, a distrbution over classes is learned for a query point $q$, and is given as 

\begin{equation}
p_{\phi}(y=k | \mathbf{q})=\frac{\exp \left(-d\left(f_{\phi}(\mathbf{q}), \mathbf{a}_{k}\right)\right)}{\sum_{k^{\prime}} \exp \left(-d\left(f_{\phi}(\mathbf{q}), \mathbf{a}_{k^{\prime}}\right)\right)}
\end{equation}

At train time, we minimize negative log probability of the positive class. The training data is generated by randomly selecting a smaller subset from the training set. Then from each class, we choose a subset of samples which are considered as support points while the rest are considered as query points. The flow diagram for few shot learning is shown in Fig.~\ref{fig:flow}.

\subsection{Projection of Capsule Net class vectors into embedding space for few shot learning}\label{sec:autocaps}

A problem in extending CapsuleNet to few shot recognition is that the final layer learns class specific embeddings. This prevents using pre-trained capsule networks or fine tuning them for different number of classes. In order to overcome this, we append an autoencoder to capsule network which takes as input the concatenated class vectors from capsule net and learns an embedding of these class vectors. To adapt it to few shot recognition, we apply the prototypical loss to these embeddings. The block diagram is shown in  Figure~\ref{fig:flow1}. 

The intuition behind using an auto-encoder is that a concatenation of class vectors represents a distribution of the input over a feature vector. However, these class vectors are obtained for specific classes. Therefore, we need an embedding which can generalize over unseen classes as well. Therefore, we choose contactive auto-encoder~\cite{rifai2011contractive}, as we want the embeddings to be similar for similar inputs yet discriminative enough for similar audios not belonging to the same speaker. This is essentially useful for the prototypical loss as the embeddings from the auto-encoder can be compared with a distance function assisting the formation of prototypes for classes with limited training samples. 
\begin{table}[]
\centering
\caption{VGG-M architecture.} 
\label{vgg_arch}
\resizebox{0.35\textwidth}{!}{%
\begin{tabular}{|l|l|l|l|l|}
\hline
\textbf{layer}  & \textbf{support} & \textbf{filt dim} & \textbf{\#filts} & \textbf{stride} \\ \hline
conv1           & 7x7              & 1                 & 96               & 2x2             \\ \hline
\textbf{mpool1} & 3x3              & -                 & -                & 2x2             \\ \hline
conv2           & 5x5              & 96                & 256              & 2x2             \\ \hline
\textbf{mpool2} & 3x3              & -                 & -                & 2x2             \\ \hline
conv3           & 3x3              & 256               & 384              & 1x1             \\ \hline
conv4           & 3x3              & 384               & 256              & 1x1             \\ \hline
conv5           & 3x3              & 256               & 256              & 1x1             \\ \hline
\textbf{mpool5} & 5x3              & -                 & -                & 3x2             \\ \hline
fc6             & 9x1              & 256               & 4096             & 1x1             \\ \hline
\textbf{apool6} & 1xn              & -                 & -                & 1x1             \\ \hline
fc7             & 1x1              & 4096              & 1024             & 1x1             \\ \hline
fc8             & 1x1              & 1024              & 1251             & 1x1             \\ \hline
\end{tabular}%
}
\end{table}

\begin{table}[]
\centering
\caption{Modified Resnet-34 Architecture}
\label{res_arch}
\resizebox{0.3\textwidth}{!}{%
\begin{tabular}{|l|l|}
\hline
\textbf{Layer name}                                                    & \textbf{Resnet-34}                                                                                       \\ \hline
\begin{tabular}[c]{@{}l@{}}conv1\\ pool1\end{tabular}         & \begin{tabular}[c]{@{}l@{}}7x7, 64, stride 2\\ 3x3, max pool, stride 2\end{tabular}             \\ \hline
conv2\_x                                                      & $\left[ \begin{array}{l}{3 \times 3,64} \\ {3 \times 3,64}\end{array}\right] \times 3$\\ \hline
conv3\_x                                                      & $\left[ \begin{array}{l}{3 \times3,128} \\ {3 \times 3,128}\end{array}\right] \times 4$\\ \hline
conv4\_x                                                      & $\left[ \begin{array}{l}{3 \times 3,256} \\ {3 \times 3,256}\end{array}\right] \times 6$\\ \hline
conv5\_x                                                      & $\left[ \begin{array}{l}{3 \times 3,512} \\ {3 \times 3,512}\end{array}\right] \times 3$\\ \hline
\begin{tabular}[c]{@{}l@{}}fc1\\ pool time\\ fc2\end{tabular} & \begin{tabular}[c]{@{}l@{}}4x1, 512, stride 1\\ 1x10, avg pool, stride 1\\ 1x1, 50\end{tabular} \\ \hline
\end{tabular}%
}
\end{table}

\section{Experiments}\label{sec:exp}

\subsection{Datasets}
We use a subset of Voxceleb1 dataset~\cite{Nagrani17} and VCTK Corpus~\cite{vctk} for few shot speaker recognition. 
\begin{itemize}[leftmargin=*]
\item Voxceleb1: VoxCeleb1 contains over $100,000$ utterances for $1,251$ speakers extracted from YouTube videos. 

Since the dataset does not provide a standard split for few shot recognition, we follow the following methodology. From the training set, we randomly sample $5$ instances $3$ second audios for each speaker. To avoid any overlap in the training data, the sampling is performed on separate training files provided in the dataset. At test time, we randomly sample $3$ second audio from the test files.  
\item VCTK: The VCTK Corpus includes speech data uttered by $109$ native speakers of English with various accents. VCTK corpus contains clearly read speech, while VoxCeleb has more background noise and overlapping speech. The dataset does not provide a standard train and test split. Therefore, we use $70:30$ train and test split. We only use a randomly sampled $3$ second audio from each split for training and evaluation purpose.
\end{itemize}

For future benchmarking and reproducibility, we will make the train and test split of the above datasets publicly available. 
    
\subsection{Results}
\subsubsection{Speaker Identification with Deep Neural Network}~\label{subsec:sp}
In the first experiment, we analyze the relative performance of various deep networks when large amount of training data is available. In Table~\ref{res_speaker_iden_pretrain}, we show results using vanilla networks with standard training and test data provided with VoxCeleb1 dataset (not few-shot setting). We indicate the capsule net architecture without reconstruction loss as CapsuleNet-M. As capsule net requires significant computational resources for large datasets~\cite{mukhometzianov2018capsnet}, we test on a subset of VoxCeleb1 for comparing against VGG and RestNet. We use first $50$ and $200$ classes of Voxceleb1 dataset and use the standard train, val and test split given for these classes. We generate the spectrogram for audio files as explained in Section~\ref{spectrogram} and feed it into the network. It can be seen that ResNet significantly outperforms the other methods and has nearly $3$ times the number of parameters as compared to VGG-M and CapsuleNet-M. Both VGG and CapsuleNet have comparable number of parameters, however, VGG performs $6$\% better on an average. In addition to the input size of $1$x$128$x$300$, we also experimented by varying number of FFT features to $256$ and $512$, and found that while it significantly increases the number of parameters, it doesn't lead to any significant boost in performance.

\begin{table}[]
\caption{Performance of deep networks on first $50$ and $200$ classes of Voxceleb1 dataset with standard train and test split Input size is $1$x$128$x$300$ in each case. NP is number of parameters.} 
\label{res_speaker_iden_pretrain}
\resizebox{\columnwidth}{!}{%
\begin{tabular}{|c|c|c|c|c|c|c|}
\hline
\multicolumn{1}{|l|}{\multirow{2}{*}{\textbf{Architecture}}} & \multicolumn{3}{c|}{\textbf{50 Classes}}     & \multicolumn{3}{c|}{\textbf{200 Classes}}    \\ \cline{2-7} 
\multicolumn{1}{|l|}{}                                       & NP         & Top-1 Acc      & Top-5 Acc      & NP         & Top-1 Acc      & Top-5 Acc      \\ \hline
CapsuleNet-M                                                 & 8,196,864  & 67.70          & 90.99          & 16,798,464 & 47.01          & 71.71          \\ \hline
VGG-M                                                        & 8,291,634  & 76.70          & 95.34          & 8,445,384  & 58.63          & 84.17          \\ \hline
Resnet-34                                                    & 22,354,162 & \textbf{90.37} & \textbf{98.13} & 22,431,112 & \textbf{71.48} & \textbf{88.45} \\ \hline
\end{tabular}}
\end{table}

\begin{table}[]
\centering
\caption{Performance of Few shot learning approach with deep neural networks on Voxceleb1}
\label{res_few_shot_learn}
 \resizebox{0.35\textwidth}{!}{%
\begin{tabular}{|c|c|c|c|c|}
\hline
\multirow{2}{*}{\textbf{Architecture}} & \multicolumn{2}{c|}{\textbf{5-way}} & \multicolumn{2}{c|}{\textbf{20-way}} \\ \cline{2-5} 
                                       & 1 shot           & 5 shot           & 1 shot            & 5 shot           \\ \hline
CapsuleNet-MA                          & 53.62            & 82.93            & 42.08             & 64.72            \\ \hline
VGG-M                                  & 52.42            & 82.10            & 20.75             & 51.82            \\ \hline
Resnet-34                              & \textbf{79.97}   & \textbf{91.50}   & \textbf{48.09}    & \textbf{72.77}   \\ \hline
\end{tabular}}
\end{table}      

 \begin{table}
            \centering
            \caption{Performance of Speaker identification on VCTK corpus (non few-shot). NP is number of parameters}
             \label{res_speak_iden_vctk}
            \resizebox{0.35\textwidth}{!}{%
\begin{tabular}{|c|c|c|c|}
\hline
\multicolumn{1}{|c|}{\textbf{Architecture}} & \multicolumn{1}{c|}{\textbf{NP}} & \multicolumn{1}{c|}{\textbf{Top-1 Acc}} & \multicolumn{1}{c|}{\textbf{Top-5 Acc}} \\ \hline
CapsuleNet-M                                & 8,196,864                        & 91.95                                   & 98.13                                   \\ \hline
VGG-M                                       & 8,291,634                        & 95.25                                   & 99.45                                   \\ \hline
Resnet-34                                   & 22,354,162                       & \textbf{96.91}                          & \textbf{99.91}                          \\ \hline
\end{tabular}}
\end{table}

\begin{table}
 \centering
 \caption{Performance of few shot speaker identification on VCTK dataset.}
\label{res_few_shot_learn_vctk}
\resizebox{0.35\textwidth}{!}{%
\begin{tabular}{|c|c|c|c|c|}
\hline
\multirow{2}{*}{\textbf{Architecture}} & \multicolumn{2}{c|}{\textbf{5-way}} & \multicolumn{2}{c|}{\textbf{20-way}} \\ \cline{2-5} 
                                       & 1 shot           & 5 shot           & 1 shot            & 5 shot           \\ \hline
CapsuleNet-MA                                & 65.26            & 91.28            & 32.45             & 68.75           \\ \hline
VGG-M                                  & 54.08            & 84.29            & 19.66             & 54.21            \\ \hline
Resnet-34                              & \textbf{80.96}   & \textbf{96.46}   & \textbf{44.95}    & \textbf{77.11}   \\ \hline
\end{tabular}}
\end{table}

\subsubsection{Effect of Number of Training Samples per Class}~\label{sec:tr}
\begin{figure}
\centering
    \includegraphics[width=7.2cm]{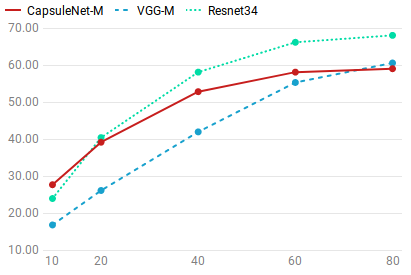}
    \caption{Test accuracy on 50 classes of Voxceleb1 for different networks trained with limited samples per class}
    \label{fig_limited_train}
\end{figure}

We now study the impact of reducing the training samples on the vanilla networks. As the previous experiment was based on training with entire training data, this setting allows us to evaluate the performance of these networks when the number of samples for each speaker reduces drastically. We vary the number of training samples per class and the results are shown in Figure~\ref{fig_limited_train}. It can be observed that with $10$ shots, the performance of all the network decreases drastically. However, CapsuleNet-M performs better than both ResNet and VGGNet. Moreover, as expected, with increasing number of samples, accuracy of all the three networks increases, with Capsule Network consistently performing better than VGG till $\sim70$ samples. This is an interesting observation and indicates that with lesser amount of samples, it is able to better exploit the structural composition of the input spectrogram. Moreover, with nearly one-third of the number of parameters as compared to ResNet, it performs close to ResNet at $10$ and $20$ shots

\subsubsection{Few Shot Learning with Prototypical Loss}
The results on few shot speaker identification are shown in     Table~\ref{res_few_shot_learn}. CapsuleNet-MA shows the results with the architecture discussed in Section \ref{sec:autocaps}. It can be observed that with prototypical loss, the performance of all the networks increases significantly. Interestingly, all the networks provide significantly better performance than just reducing the number of training samples (Section \ref{sec:tr}). Moreover, the capsule network with autoencoder outperforms VGG while performs closer to ResNet. Moreover, ResNet provides close to $80$\% accuracy with 1-shot for $5$ speakers and nearly
 $70$\% with 5-shot classification for 20 speakers. This is important as it indicates that even with heavily constrained settings one can identify the speaker with high confidence. 
\vspace{-0.7em}
\subsubsection{Generalization Ability of Networks}
To show the generalization ability of networks, first we use the trained models on $50$ classes of Voxceleb1 and fine-tune them on first 50 classes of VCTK corpus for speaker identification task. We report the accuracy in Table~\ref{res_speak_iden_vctk}. We also used pre-trained models on Voxceleb1 for few shot learning task and use them to perform few shot classification on VCTK Corpus without fine-tuning them and report these results in Table~\ref{res_few_shot_learn_vctk}. It can be observed that, with pretrained networks the method is also able to generalize entirely to an unseen class with samples collected using entirely different criteria. It can be observed that from Table~\ref{res_speak_iden_vctk}, that by just fine tuning the networks, ResNet achieves a top-$5$ accuracy of $\sim99\%$ while VGG and CapsuleNet follow similar trends with their accuracies being slightly behind ResNet. However, in case of few shot recognition (Table~\ref{res_few_shot_learn_vctk}, we notice significant drop in the performance of VGG while CapsuleNet-MA performs $\sim10$\% better than VGG on an average.  

\section{Conclusion}\label{sec:conc}
We have shown effectiveness of few shot learning approaches to speaker identification. We also demonstrated that with deep neural networks, one can identify speakers with high confidence with ResNet outperforming other techniques by a significant margin. However, number of parameters for ResNet were comparatively high. On the other hand, CapsuleNet performed better on less amount of data but applying it for few shot recognition is not trivial. Therefore, we proposed an extension of Capsule Network by learning a generalized embedding using a contractive auto-encoder. The computed embeddings are used for learning a prototype emebedding using prototypical loss. We believe that this work will accelerate use of CNN for practical implementations as well as catalyze the research on Capsule Network for making it more efficient on large scale data. 

\bibliographystyle{IEEEtran}
\bibliography{mybib}

\end{document}